\documentclass[aip,apl,amsmath,floatfix,reprint]{revtex4-1}
\usepackage{graphicx,psfrag,mathptmx}

\begin{document}

\title{Shadow evaporation of epitaxial Al/Al$_2$O$_3$/Al tunnel junctions on sapphire utilizing an inorganic bilayer mask}
\thanks{This work was sponsored by the Assistant Secretary of Defense for Research \& Engineering under Air Force Contract No.~FA8721-05-C-0002.  Opinions, interpretations, conclusions and recommendations are those of the authors and are not necessarily endorsed by the United States Government.}
\author{Paul B. Welander}\email{welander@ll.mit.edu}
\author{Vladimir Bolkhovsky}\author{Terence J. Weir}\author{Mark A. Gouker}\author{William D. Oliver}
\affiliation{Lincoln Laboratory, Massachusetts Institute of Technology, Lexington, MA 02420}
\date{March 27, 2012}
\begin{abstract}
This letter describes a new inorganic shadow mask that has been employed for the evaporation of all-epitaxial Al/Al$_2$O$_3$/Al superconducting tunnel junctions.  Organic resists that are commonly used for shadow masks and other lift-off processes are not compatible with ultra-high vacuum deposition systems, and they can break down at even moderately elevated temperatures.  The inorganic mask described herein does not suffer these same shortcomings.  It was fabricated from a Ge/Nb bilayer, comprising suspended Nb bridges supported by an undercut Ge sacrificial layer.  Utilizing such a bilayer mask on C-plane sapphire, the growth of epitaxial Al tunnel junctions was achieved using molecular beam epitaxy.  Crystalline Al$_2$O$_3$ was grown diffusively at 300~$^{\circ}$C in a molecular oxygen background of 2.0~$\mu$torr, while amorphous oxide was grown at room temperature and 25~mtorr.  A variety of analysis techniques were employed to evaluate the materials, and tunnel junction current-voltage characteristics were measured at millikelvin temperatures.
\end{abstract}
\maketitle

A common technique for the fabrication of Josephson junctions involves double-angle shadow evaporation of Al through an offset mask, with the tunnel barrier formed by the diffusive oxidation of the Al base layer. \cite{Dolan1977,Fulton1987}  Shadow evaporation has been the most successful fabrication approach to date for making long-lived, high-coherence superconducting quantum bits (or qubits). \cite{Bylander2011,Steffen2010, DiCarlo2010,Paik2011}  An alternative technique for tunnel junction fabrication involves the selective etching of whole-wafer superconductor/insulator/superconductor (SIS) trilayers. \cite{Gurvitch1983,Ketchen1991}  Spectroscopy measurements on qubits fabricated in this manner have shown that epitaxial Al$_2$O$_3$ tunnel barriers have fewer two-level fluctuators (a known source of decoherence) than diffused, amorphous oxides.\cite{Oh2006}  Given the respective successes of these techniques, combining the two would seem ideal.  However, the conventional organic bilayers used for shadow masks are incompatible with the ultra-high vacuum (UHV) environment and the high substrate temperatures required for epitaxial film growth.

To address this challenge we have developed an inorganic bilayer mask, and used it to combine epitaxial trilayer growth with shadow evaporation for the fabrication of crystalline Josephson junctions.  Inorganic masks have been utilized before by a few different groups, but our interest was developing one consistent with our existing qubit fabrication process. \cite{Gustavsson2012}  Those developed by Howard \cite{Howard1978} used Cu for a sacrificial lift-off layer, and Cu is incompatible with our CMOS fabrication facility because of its degradation of minority carrier lifetimes in Si. \cite{Istratov1997}  Offset masks used by van Wees \textit{et al.} \cite{vanWees1985} employed Al as the lift-off layer, which doesn't permit the fabrication of Al tunnel junctions.  Finally, Hoss \textit{et al.} \cite{Hoss1999} utilized SiO$_2$ for the lift-off layer, a dielectric material incorporated into our qubit fabrication process for wiring cross-overs.  Given these limitations, we sought to develop a CMOS-compatible process for Al/Al$_2$O$_3$/Al tunnel junctions.  In addition to being compatible with our qubit fabrication process, we also desired to resolve suspended bridges down to 100 nm wide and up to 2.5~$\mu$m long.

\begin{figure}[b]
  \includegraphics[width=3.375in]{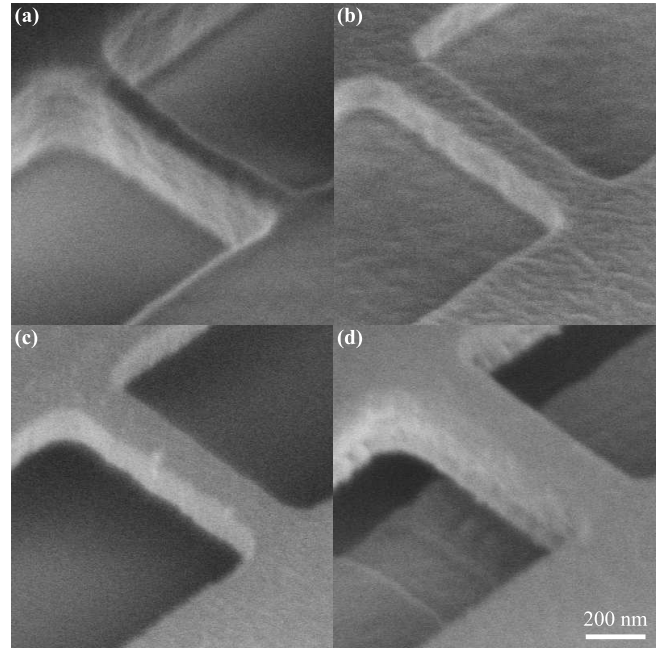}
  \caption{The process for making an inorganic Ge/Nb bilayer mask for Al tunnel junction shadow evaporation.  (a) The bridge pattern is first defined in an ARC/UV5 resist bilayer using 248-nm photolithography.  (b) The top metal (Nb) is etched using Fl-based RIE and the resist layers are stripped.  (c) The sacrificial layer (Ge) is undercut with H$_2$O$_2$.  (d) Al/Al$_2$O$_3$/Al shadow evaporation is performed, followed by lift-off in H$_2$O$_2$.}
  \label{Figure1}
\end{figure}

Our inorganic shadow masks were fabricated from Ge/Nb bilayers in the Microelectronics Laboratory at MIT Lincoln Laboratory (see Fig.~\ref{Figure1}).  The sacrificial Ge layer was about 500~nm thick and the Nb bridge layer was about 100~nm thick.  To pattern the top Nb layer, we used 248~nm deep-ultraviolet (DUV) photolithography with UV5 resist and an anti-reflective coating (ARC), followed by F-based reactive ion etching (RIE).  After stripping the organic resist layer from the wafer, the final step in preparing the shadow mask was to undercut the Ge layer with a H$_2$O$_2$ wet-etch such that the Nb overhang was roughly equal to the Ge thickness.  Fig.~\ref{Figure1} shows this process, with which we can routinely make bridges down to 150~nm wide and up to 2.5~$\mu$m long.  (We have employed e-beam lithography to define even narrower bridges, but that approach was not employed in the work discussed here.)

\begin{figure}[t]
  \includegraphics[width=3.375in]{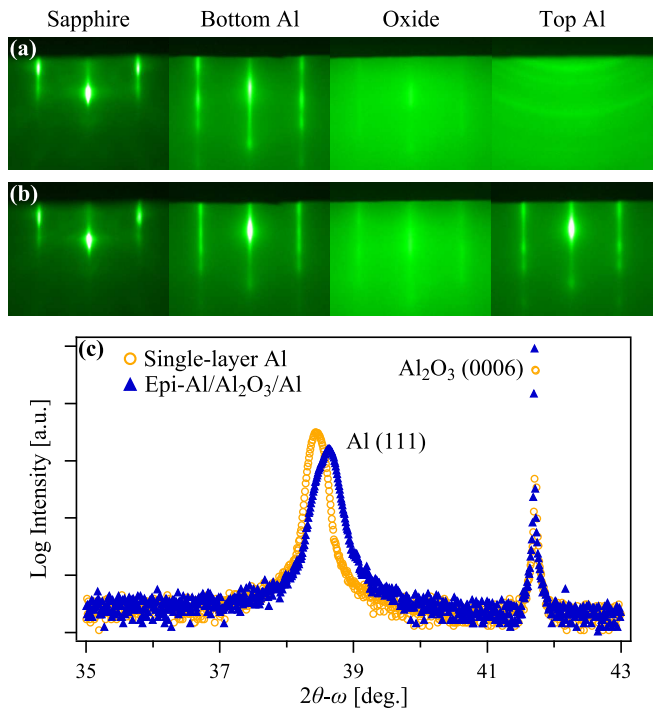}
  \caption{(Color online) Series of RHEED images from shadow-evaporated Al/Al$_2$O$_3$/Al tunnel junctions with amorphous (a) and crystalline (b) oxide barriers.  In both cases the bottom Al layer is (111)-oriented and RHEED from the Al$_2$O$_3$ layer is quite diffuse.  The distinguishing characteristic is the crystallinity of the top Al layer: growth on amorphous oxide yields polycrystalline Al, while the crystalline oxide gives rise to (111)-oriented Al.  The bottom panel (c) shows an XRD 2$\theta$-$\omega$ scan comparing single layer Al with an all-epitaxial Al/Al$_2$O$_3$/Al film.  The latter exhibits a shift of about 0.4\% toward smaller planar spacing.}
  \label{Figure2}
\end{figure}

For standard, non-epitaxial, Al shadow evaporation processes, we fabricated inorganic masks on both bare Si and Si with thermal oxide.  For epitaxial Al growth, we utilized C-plane sapphire -- $\alpha$-Al$_2$O$_3$ (0001) -- for several reasons: 1.)~it does not react with H$_2$O$_2$; 2.)~it is easily cleaned with solvents (the Ge/Nb bilayer is not chemically attacked); 3.)~the surface remains crystalline when exposed to air; and 4.)~Al (111) films have been demonstrated on C-plane sapphire. \cite{Vermeersch1995,Medlin1997,Dehm2002}  In the center of the sapphire substrate, an area of about 5~cm$^2$ was flood-exposed and completely stripped of the mask layers -- this enabled \textit{in situ} reflection high-energy electron diffraction (RHEED) measurements before, during, and after film growth.  Crystalline Al growth was achieved using a molecular beam epitaxy (MBE) system with a base pressure less than $1\times10^{-10}$~torr and an effusion cell for Al deposition.  The Al cell was mounted at a fixed angle of 45$^{\circ}$, and RHEED was used to adjust the wafer's azimuthal angle such that the mask pattern was aligned to the Al cell position.  Epitaxial Al (111) films were achieved through room-temperature deposition without any pretreatment of the wafer in UHV.

Amorphous oxide tunnel barriers were formed by oxidizing the Al (111) surface in the MBE load-lock at an O$_2$ pressure of 25~mtorr for 30~min.  Crystalline oxide tunnel barriers were formed in the growth chamber by diffusive oxidation of the Al (111) surface at high temperatures. \cite{Jeurgens2008,Reichel2008}  Several growth conditions were evaluated using both RHEED to measure the surface crystallinity, and x-ray photoelectron spectroscopy (XPS) to measure oxide thickness.  Comparing the XPS spectra of crystalline oxide layers with that of an amorphous oxide formed under the conditions above, the crystalline oxide process that gave the desired thickness had a wafer temperature of 300~$^{\circ}$C, an O$_2$ pressure of 2.0~$\mu$torr, and a duration of 5~hours.  After forming the oxide barrier layer and depositing the top Al electrode layer, the Ge/Nb mask layers were lifted off in H$_2$O$_2$, leaving behind the Al/Al$_2$O$_3$/Al overlap Josephson junction.

\begin{figure}[t]
  \includegraphics[width=3.375in]{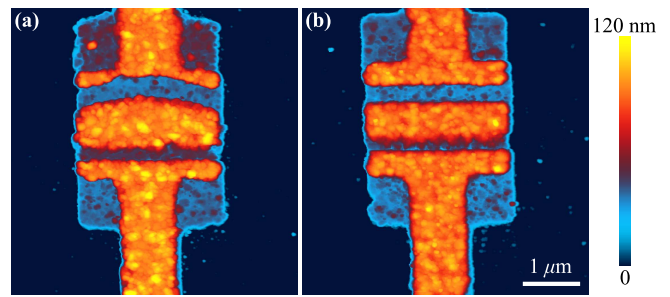}
  \caption{(Color online) AFM scans of two shadow-evaporated Al/Al$_2$O$_3$/Al tunnel junctions following lift-off of the inorganic mask. In both cases the bridge width and length measured 200~nm and 2.5~$\mu$m, respectively.  For the crystalline oxide case (a), the heating of the wafer to 300~$^{\circ}$C during oxidation caused the bridge to bow up slightly, resulting in the curved shadow produced during the second Al evaporation step.  Compare with the amorphous oxide case (b), where the bottom Al layer was oxidized at ambient temperature.}
  \label{Figure3}
\end{figure}

RHEED measurements of the two junction varieties (with amorphous and crystalline barriers) are shown in Fig.~\ref{Figure2}.  In both cases the Al~(111) base layer grows epitaxially on C-plane sapphire with the orientational relationship Al~$[11\bar{2}] \parallel$~$\alpha$-Al$_2$O$_3$~$[11\bar{2}0]$.  An amorphous oxide layer leads, as expected, to a polycrystalline top Al layer, indicated by the faint rings observed in the RHEED pattern (see Fig.~\ref{Figure2}a).  On the other hand, the crystalline oxide gives rise to an epitaxial Al (111) top layer with a RHEED pattern nearly indistinguishable from the bottom Al layer.  The oxide layer itself has a very diffuse diffraction pattern, quite different from the sapphire substrate, so it is somewhat surprising that the two Al layers look so similar from a RHEED perspective.  We speculate that the polymorphic nature of Al$_2$O$_3$ \cite{Wefers1987,Levin1998} gives rise to well-ordered, close-packed oxygen planes with an aluminum sublattice that is either disordered or has order only over limited length scales, thus producing the diffuse RHEED pattern.  This explanation is plausible because all of the known sapphire polymorphs are thermodynamically stable at temperatures up to 300 $^\circ$C and beyond.  In addition, many polymorphs (eg. $\gamma$, $\eta$, $\delta$, and $\theta$) are less dense and have larger lattice parameters.  We note this because the atomic spacing of the Al (111) surface lattice is about 4\% larger than that of $\alpha$-Al$_2$O$_3$ (0001) -- 2.864 vs.~2.748 \AA, respectively -- and this misfit may be partially accommodated through the growth of one or more polymorphs.

X-ray diffraction (XRD) measurements of the Al trilayers after lift-off confirmed the orientation of the Al layers, in both the surface-normal and in-plane directions.  Like other close packed metals, Al epitaxy in the (111) growth orientation exhibited stacking faults, as evidenced by the six-fold symmetry observed in $\phi$-scans (not shown).  The crystalline Al trilayer also showed a slightly wider rocking curve ($\omega$-scan) than did a single layer Al film -- 0.21$^{\circ}$ versus 0.15$^{\circ}$.  Finally, radial scans of trilayer and single layer Bragg peaks showed a slight shift of the trilayer Al (111) peak to a higher $2\theta$-$\omega$ value (see Fig.~\ref{Figure2}c).  While single-layer Al was found to have a (111) planar spacing, $d_{\textrm{(111)}}=2.340$~\AA, very close to that of bulk Al (2.338~\AA), the crystalline trilayer exhibited a smaller spacing of 2.330~\AA.  The cause for this contraction of the Al trilayer lattice is not known.

\begin{figure}[t]
  \includegraphics[width=3.000in]{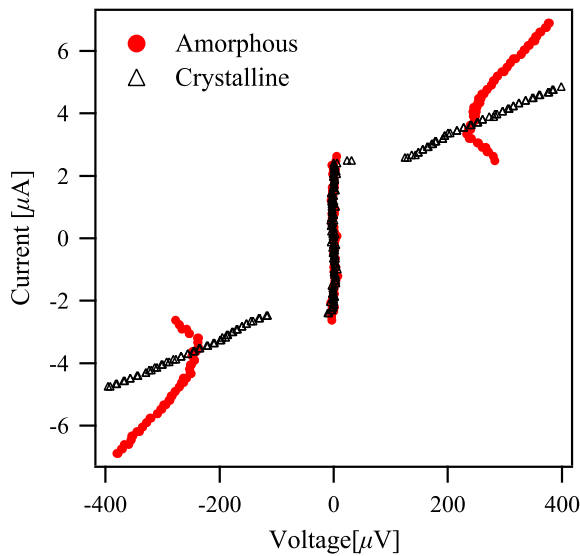}
  \caption{(Color online) Current-voltage characteristics for two Al tunnel junctions: one with an amorphous oxide layer ({\large$\bullet$}) and one with a crystalline oxide layer ({\scriptsize$\triangle$}).  Both junctions were formed with a bridge 160~nm wide and 1000~nm long, yielding an overlap of about 650~nm.  The critical current density in both cases is about 4~$\mu$A/$\mu$m$^2$.  The amorphous barrier shows hysteresis and a gap voltage of about 240~mV while the crystalline barrier is electrically leaking and superconductivity in the Al base layer appears suppressed.}
  \label{Figure4}
\end{figure}

After lift-off the devices were also scanned using atomic force microscopy (AFM) both to measure the Al films' thickness at various locations across the wafer, and to measure the overlap achieved for a variety of bridge widths.  We found that across a 150 mm-diameter wafer the thickness of each Al layer varied about $\pm20\%$, and that for a 200 nm-wide bridge the overlap of the two Al layers was about 600~nm.  This overlap length is consistent with the estimate $(2d-w-t-a)$ one can make for 45$^{\circ}$ evaporation, where $d$ is the thickness of the sacrificial Ge layer, $w$ and $t$ are the bridge width and thickness, respectively, and $a$ is the thickness of the Al base layer.  We also observed in the AFM scans that longer bridges tended to bow upward during the crystalline oxide growth process, which led to an increased overlap in those devices.  This is shown in Fig.~\ref{Figure3} for two junctions formed with a bridge 200~nm wide by 2.5~$\mu$m long -- one with crystalline oxide formed at 300~$^{\circ}$C (Fig.~\ref{Figure3}a) and one with an amorphous oxide formed at room temperature (Fig.~\ref{Figure3}b).  Eliminating this bow, the subject of future work, may be achieved through better stress control in the Nb and/or Ge layers during deposition, or by thermal treatment of the bilayer prior to shadow-mask patterning.

Electrical measurements of tunnel junction current-voltage characteristics ($I$-$V$'s) were performed in a dilution refrigerator operating below 20~mK.  Typical $I$-$V$'s for both crystalline and amorphous oxide barriers are shown in Fig.~\ref{Figure4}.  Junctions with an amorphous barrier exhibited the common, hysteretic curve of a SIS tunnel junction, with features at the gap voltage due to self-heating.  On the other hand, crystalline tunnel barriers displayed no such hysteresis and had a gap roughly half that observed with the amorphous barrier.  While enabling epitaxial growth from bottom to top, the crystalline oxide does not form an electrically insulating film and is most likely littered with electrical pinholes.  It is also plausible that the process for forming crystalline Al$_2$O$_3$ on Al diffuses a substantial amount of oxygen into the bulk of the metal base layer, to the point where superconductivity is suppressed to a large degree.

In summary, we have developed a new inorganic bilayer mask and utilized it to grow epitaxial Al/Al$_2$O$_3$/Al Josephson junctions.  The shadow mask was fabricated from a Ge/Nb bilayer deposited on C-plane sapphire, and while suspended bridges did show some bowing due to thermal stress, the mask held up well under the growth conditions studied.  Both electron and x-ray diffraction measurements on epitaxial Al trilayers indicated very good crystalline quality was achieved.  However, the high-temperature diffusive oxidation process employed yielded leaky tunnel barriers and may have suppressed superconductivity in the base Al layer.  An alternative growth technique such as co-deposition \cite{Welander2010} may be more successful in this regard, and is a subject for future research.

\vspace{10 pt}

The authors would like to thank Joseph Powers, Peter Murphy and Jacob Zwart for technical assistance, and Daniel Calawa for performing the XRD measurements.  This work was sponsored by the Assistant Secretary of Defense for Research \& Engineering under Air Force Contract No.~FA8721-05-C-0002.  Opinions, interpretations, conclusions and recommendations are those of the authors and are not necessarily endorsed by the United States Government.

%\bibliography{epi_shadevap}

%Merlin.mbs v4.21 2009-07-09.
%

\end{document}